\documentclass[aip,apl,reprint]{revtex4-1}

\usepackage{graphicx}
\usepackage{dcolumn}
\usepackage{bm}
\usepackage{url}
\usepackage{color}

\usepackage{amssymb} 

\usepackage{units}

\begin{document}


\title{On-and-off chip cooling of a Coulomb blockade thermometer down to 2.8\,mK}

\author{M.~Palma}\altaffiliation{These authors contributed equally to this work.}
\affiliation{Department of Physics, University of Basel, Klingelbergstrasse 82, CH-4056 Basel, Switzerland}

\author{C.~P.~Scheller}\altaffiliation{These authors contributed equally to this work.}
\affiliation{Department of Physics, University of Basel, Klingelbergstrasse 82, CH-4056 Basel, Switzerland}

\author{D.~Maradan}
\affiliation{Department of Physics, University of Basel, Klingelbergstrasse 82, CH-4056 Basel, Switzerland}
\affiliation{Physikalisch-Technische Bundesanstalt~(PTB), Bundesallee 100, 38116 Braunschweig, Germany}

\author{A.~V.~Feshchenko}
\affiliation{Low Temperature Laboratory, Department of Applied Physics, Aalto University,P.O. Box 13500, FI-00076 AALTO, Finland}

\author{M.~Meschke}
\affiliation{Low Temperature Laboratory, Department of Applied Physics, Aalto University,P.O. Box 13500, FI-00076 AALTO, Finland}


\author{D.~M.~Zumb\"{u}hl}
\email[]{dominik.zumbuhl@unibas.ch}
\affiliation{Department of Physics, University of Basel, Klingelbergstrasse 82, CH-4056 Basel, Switzerland}

\date{\today}

\begin{abstract}
Cooling nanoelectronic devices below 10\,mK is a great challenge since thermal conductivities become very small, thus creating a pronounced sensitivity to heat leaks. Here, we overcome these difficulties by using adiabatic demagnetization of \emph{both} the electronic leads \emph{and} the large metallic islands of a Coulomb blockade thermometer. This reduces the external heat leak through the leads and also provides on-chip refrigeration, together cooling the thermometer down to $2.8\pm0.1\,$mK. We present a thermal model which gives a good qualitative account and suggests that the main limitation is heating due to pulse tube vibrations. With better decoupling, temperatures below $1\,$mK should be within reach, thus opening the door for $\mu$K nanoelectronics.
\end{abstract}

\maketitle 

\par Reaching ultralow temperatures in electronic transport experiments can be key to novel quantum states of matter such as helical nuclear spin phases~\cite{Simon2007,Simon2008,Scheller2014a}, full nuclear spin polarization\cite{Chesi2008}, quantum Hall ferromagnets~\cite{Chesi2008} or fragile fractional quantum Hall states~\cite{Pan2015,Samkharadze2015}. In addition, the coherence of semiconductor and superconducting qubits~\cite{Hanson2007,Clarke2008,Devoret2013} as well as hybrid Majorana devices~\cite{Lutchyn2010,Oreg2010,Alicea2010,Mourik2012} could benefit from lower temperatures. With this motivation in mind, we built a parallel network of nuclear refrigerators~\cite{Clark2010} to adapt the very well established technique of Adiabatic Nuclear Demagnetization (AND)~\cite{Pickett1988,Pickett2000,Pobell2007} for electronic transport experiments. In this approach, the concept is to cool a nanoelectronic device directly through the electronic leads, which remain effective thermal conductors also below 1\,mK. Each wire is cooled by its own, separate nuclear refrigerator in form of a large Cu plate. However, despite recent progress~\cite{Casparis2012,batey_2013,Maradan2014,Feshchenko2015,Bradley2016,Iftikhar2016,Bradley2017,Palma2017}, it remains very challenging to cool nanostructures even below 10\,mK. Due to reduced thermal coupling, these samples are extremely susceptible to heat leaks such as vibrations\cite{Palma2017}, microwave radiation\cite{Saira2012,Zorin1995}, heat release\cite{Pobell2007} and electronic noise\cite{Maradan2014}.

\begin{figure}
	\centering
	\includegraphics[width=1\columnwidth]{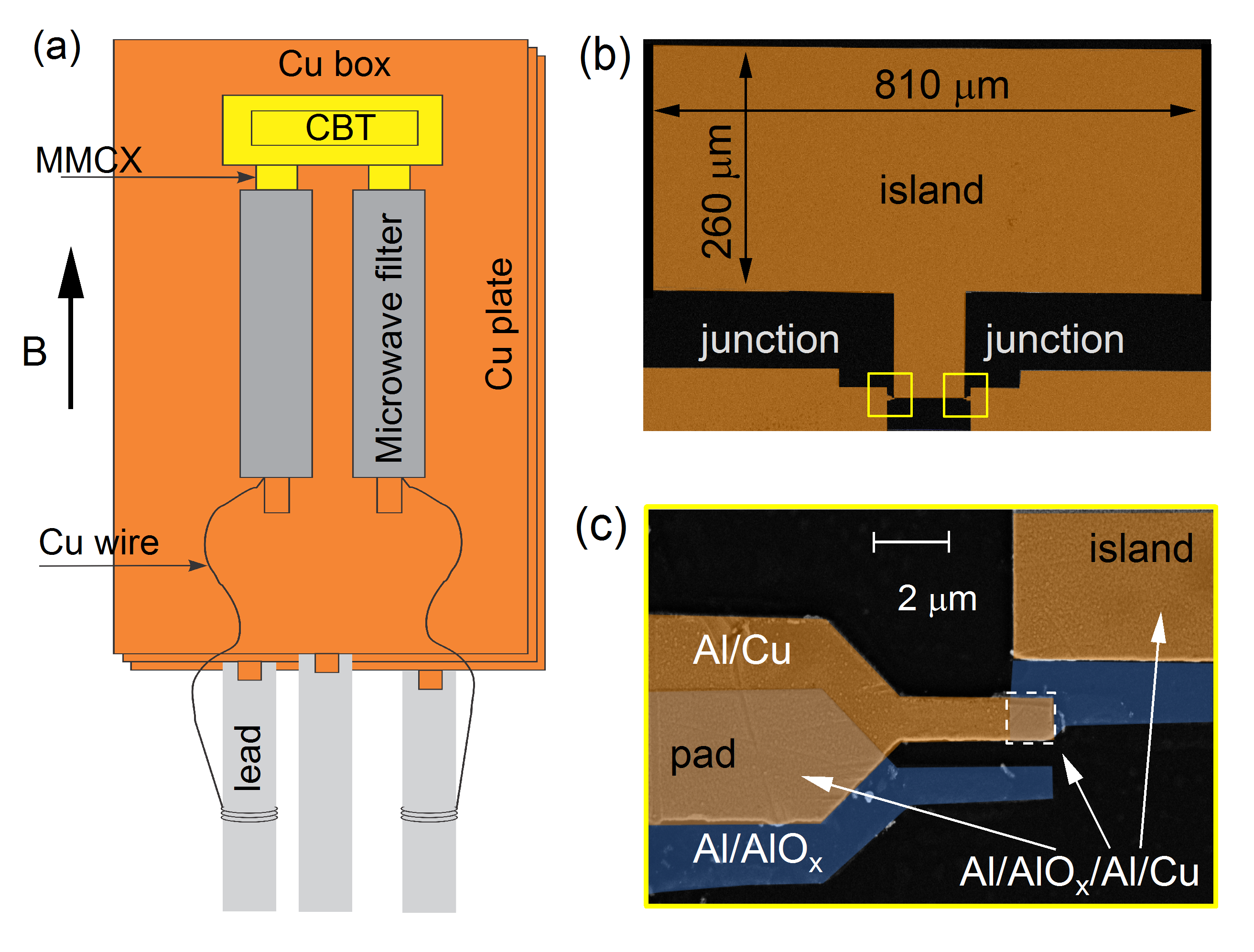}\vspace{-2.5mm}
	\caption{ (a) Schematic with CBT enclosed in a copper box (yellow), connected to Ag-epoxy microwave filters (grey), and glued onto a Cu plate (orange) with Ag-epoxy. (b) Electron micrograph (false color) of the CBT island (volume $\approx~42,000~\mathrm{\mu m^{3}}$), with tunnel junctions (inside yellow rectangles) to adjacent $\mathrm{Al/AlO_x/Al/Cu}$ pads. The brown sections comprise very large areas, giving very low resistances of these large junctions. (c) Zoom-in of a tunnel junction from (b), showing the overlap (white rectangle) between the top layer (Al/Cu, brown) and the bottom layer (Al/AlOx, blue).}\vspace{-3mm}
	\label{fig:CBT_scheme}
\end{figure}

Metallic Coulomb blockade thermometers (CBTs) have been established as precise and reliable electronic thermometers \cite{Pekola1994,Meschke2004,Casparis2012}, operating down to 10\,mK and slightly below \cite{Scheller2014,Feshchenko2015,Bradley2016,Bradley2017}. These typically consist of linear arrays of $\mathrm{Al/AlO_x/Al}$ tunnel junctions with metallic islands in-between, consisting mainly of copper, see Fig.\,\ref{fig:CBT_scheme}. The array divides the electronic noise per island by the number of junctions in series. This makes them less susceptible to electronic noise, but thermal conduction via Wiedemann-Franz cooling is not very effective through a series of resistive tunnel junctions. For this reason, the islands were enlarged into giant cooling fins \cite{Meschke2004} providing a huge volume for effective electron-phonon coupling and cooling through the substrate. At low temperatures, however, this eventually fails due to the very strong $T^5$ temperature dependence of the electron phonon coupling.

For a \emph{single} tunnel junction\cite{Nahum1993,Feshchenko2015}, on the other hand, two low-resistance reservoirs adjacent to the junction can be cooled efficiently by electronic leads making contact to nuclear refrigerators. Thus, in a noisy environment, on-chip nuclear demagnetization of the large fins offers itself as an elegant solution for a CBT array, providing local, in-situ cooling without having to go through potentially ineffective tunnel barriers or an insulating substrate. The large volume of the metallic islands now is taken advantage of as the spin reservoir of an AND refrigerator. At sufficiently low temperature, the electron-phonon coupling becomes so weak that the islands decouple thermally from the substrate, thus giving a low heat leak, as desired for efficient cooling. Previously, AND was applied to AlMn based single electron transistors~\cite{Ciccarelli2016} and CBTs with electroplated islands~\cite{Bradley2016,Bradley2017}, in both cases reducing the electronic temperature by roughly a factor of two.

\par In this Letter, we perform AND in \emph{both} the Cu plates in the leads \emph{and} the massive CBT islands, thus combining direct on-chip cooling with a reduced external heat leak emanating from the leads. We obtain significantly improved cooling, lowering the electronic temperature by a factor of 8.6, from $\sim$24\,mK down to 2.8$\pm$\,0.1\,mK, thereby further reducing the lowest reported electronic temperature in a solid state device \cite{Bradley2017}. We present a simple model giving a qualitative account of the cooling cycle. The performance is limited by a heat leak caused by the pluse-tube vibrations. With improved decoupling, the micro-Kelvin regime in nanoelectronics should be within reach.

\par The present experiment is performed on a cryo-free platform~\cite{bluefors,Palma2017}, where each of the 16 leads is equipped with its own nuclear refrigerator (see Fig.\,\ref{fig:CBT_scheme}), consisting of two moles of copper. The system allows for cooling of the Cu pieces down to $T_{\rm{Cu}}$=~150\,$\mu$K~\cite{Palma2017}, see supplementary for details\cite{SOM}. To integrate the CBT, it is placed inside a Cu box, see Fig.~\ref{fig:CBT_scheme}(a), which will also be demagnetized and further shields the CBT from high frequency radiation, arising e.g. from higher temperature stages of the refrigerator. The Cu box is equipped with coaxial MMCX connectors, and connected directly to additional Ag-epoxy microwave filters~\cite{Scheller2014}. The microwave filters used here are made from rather thick copper wire (0.35\,mm instead of 0.1\,mm) and contain a thicker and longer copper core, thus facilitating direct demagnetization of the filters themselves. The Cu box and the microwave filters are glued onto a Cu nuclear refrigerator with conductive Ag-epoxy in order to ensure good thermalization. The CBT consists of an array of 16 large metallic islands with $1\,\mathrm{\mu}$m$^2$ $\mathrm{AlO_x}$ tunnel barriers in-between, as shown in Fig.\,\ref{fig:CBT_scheme}(b,c). This gives a device resistance of $150\,\mathrm{k\Omega}$ and provides a rather small charging energy $E_{c}\approx6.5\,$mK, thus allowing accurate thermometry down to $\approx2\,$mK~\cite{Feshchenko2013}. The islands are of almost macroscopic size (Cu layer $\approx 810\cdot260\cdot0.2\,\mu\rm m^3$), making available a large reservoir of nuclear spins for demagnetization\cite{SOM}.

\begin{figure}[t]
	\centering
	\includegraphics[width=1\columnwidth]{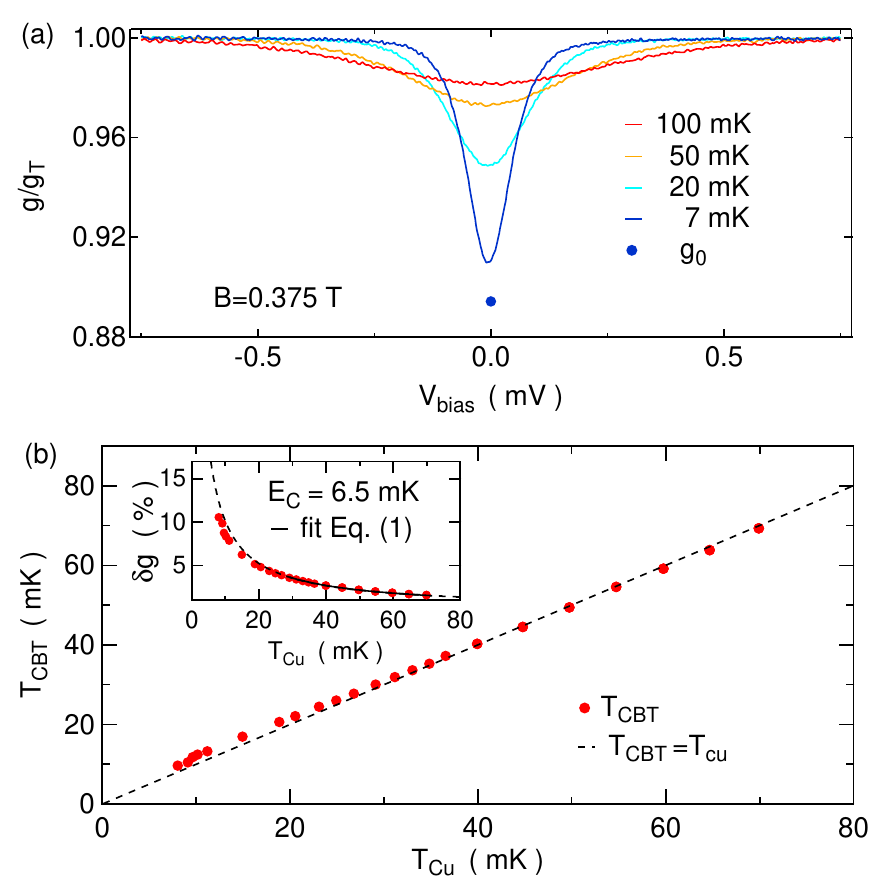}\vspace{-2.5mm}
	\caption{(a) Bias dependence of the differential conductance $g$ normalized with the high bias conductance $g_T$, shown for various Cu plate temperatures $T_{\rm{Cu}}$. An in-plane field $B=0.375\,$T drives the Al thin films normal. The zero bias conductance (dark blue circle, measured after equilibrating at $T_{\rm{Cu}}=7\,$mK) remains clearly below the bias sweep (dark blue curve). (b) CBT temperature $T_{\rm{CBT}}$ versus $T_{\rm{Cu}}$. The diagonal dashed line indicates ideal thermalization $T_{\rm{CBT}}=T_{\rm{Cu}}$. The inset shows the normalized zero bias conductance dip $\delta g$ as a function of $T_{\rm{Cu}}$. A fit using Eq.\,\ref{eq:calibration} is done over the high temperature, well thermalized regime $T_{\rm{Cu}}\geq 30\,$mK (solid black curve) and delivers the charging energy $E_c$ of the device as the only fit parameter. The dashed curve indicates the low temperature extension below 30\,mK of the fit with the same $E_c$. }\vspace{-2mm}
	\label{fig:differential}
\end{figure}	

\par In order to extract the electronic temperature of the device, we measure the 2-wire differential conductance $g$ of the CBT as a function of the bias voltage $V_{\rm{bias}}$ by means of a standard low-frequency lock-in technique using a few $\mu V$ of AC excitation. Figure~\ref{fig:differential}(a) shows typical conductance traces, measured at various refrigerator temperatures, as indicated. Due to Coulomb blockade effects\cite{Ingold}, the zero bias conductance $g_0$ is suppressed below its asymptotic, large bias value $g_{\rm{T}}$. Both width and depth of the conductance dip are commonly used for thermometry\cite{Meschke2004,Casparis2012}. While fitting the full conductance trace allows one to use the CBT as a primary thermometer, using the depth of the zero bias dip requires pre-calibration and thus can be used only as a secondary thermometer. However, the primary mode is prone to overheating due to the large applied DC bias. This effect becomes particularly important at the lowest temperatures, illustrated in Fig.\,\ref{fig:differential}(a). The conductance $g_0$ measured while permanently staying at zero bias \footnote{This requires careful input voltage drift stabilization as provided by our home built IV converter, particularly below 10~mK \cite{BaselElectronics}} is clearly lower in conductance (dark blue marker) -- and thus also lower in temperature -- than the one obtained from a bias sweep at the same refrigerator temperature (dark blue trace), see also SOM\cite{SOM}. Therefore, the CBT is used in secondary mode here for the rest of this work~\cite{Meschke2004,Casparis2012,Scheller2014}.

The normalized Coulomb blockade zero bias dip $\delta g=1-g_{0}/g_{T}$ is given by the third order expansion~\cite{Meschke2004}
\begin{equation}
\label{eq:calibration}
 \delta g=u/6-u^2/60+ u^3/630\,,
\end{equation}
\noindent for sufficiently small $u$ (see below), where $u=E_{c}/k_{B}T_{\rm{CBT}}$, and $k_{B}$ is the Boltzmann constant. The inset of Fig.\,\ref{fig:differential}(b) shows the measured $\delta g$ as a function of temperature $T_{\rm{Cu}}$ (red circles) along with a fit (solid black) performed in the high temperature regime above $30\,$mK using Eq.\,\ref{eq:calibration}. The fit delivers the charging energy $E_{c}$\,=\,6.5$\pm0.1\,$mK, which now allows us to convert the measured $\delta g$ to $T_{\rm{CBT}}$ for the whole temperature regime, thus providing the calibration of the secondary thermometer. The fit agrees very well with the data for high temperatures, but weak overheating is seen at lower temperatures $T_{\rm{CBT}}\lesssim$~30~mK, see Fig.~\ref{fig:differential}b inset.

The CBT calibration curve Eq.~\ref{eq:calibration} is becoming more benign at the lowest temperatures, where a much larger change in $\delta g$ is required for a given change in $T_{\rm{CBT}}$ compared to high temperatures. Thus, a small deviation in $\delta g$ has a very small effect on the temperature reading and the calibration curve becomes more accurate at the lowest temperatures. The validity of Eq.\,\ref{eq:calibration} was investigated in detail in Ref.~\onlinecite{Feshchenko2013}, showing that an accurate temperature reading to within $\sim$10\,\% error or less is obtained as long as $u\lesssim 3$. In practice, this means that the extracted CBT temperature, shown in Fig.\,\ref{fig:differential}(b) as a function of Cu plate temperature $T_{\rm{Cu}}$, would be quite precise down to roughly 2\,mK.

\par In the second part of this Letter, we address the simultaneous adiabatic nuclear demagnetization (AND) of the CBT and its leads. AND is a single shot technique comprising three stages. First, magnetization of the nuclear spins at large initial magnetic field $B_{\rm{i}}=$~9~T is performed. Then, the actual demagnetization is carried out down to a final field $B_{\rm{f}}$ of 0.375\,T. Finally, the warm up stage commences, where a heat load is warming up the system until the magnetization is exhausted. Here, $B_{\rm{f}}$ is limited by the critical field required to break superconductivity of the Al thin film in the CBT.

\begin{figure}[!ht]
	\centering
	\includegraphics[width=1\columnwidth]{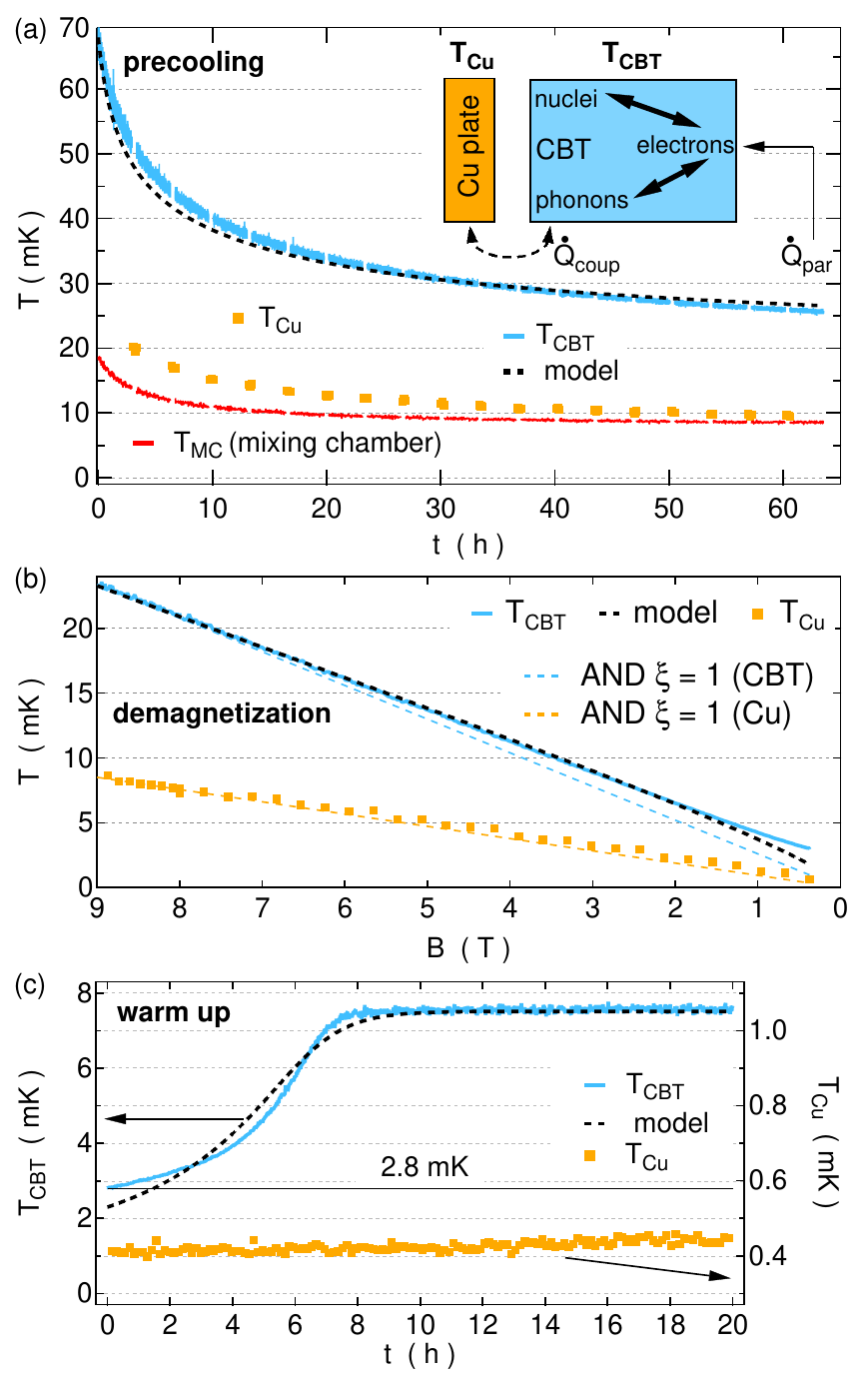}\vspace{-2.5mm}
	\caption{(a) Various temperatures as a function of the precooling time, as labeled. The inset shows a schematic of the thermal model. (b) Evolution of various temperatures during the AND process. Blue and orange dashed line indicate ideal cooling of CBT and Cu plates, respectively. (c) Warm up curves for various thermometers. The model is shown as black dashed curve for all panels.}\vspace{-3mm}
	\label{fig:cbt_AND}
\end{figure}

Upon increasing the magnetic field to $9\,$T, a large amount of heat due to the magnetization of the nuclear spins has to be drained into the mixing chamber in order to polarize the nuclear spins during precooling, depicted in Fig.~\ref{fig:cbt_AND}(a). The heat switches of the parallel refrigerator network are therefore driven normal to obtain a strong thermal link to the mixing chamber. While $T_{\rm{Cu}}$ (orange markers, Fig.\,\ref{fig:cbt_AND}(a)), eventually closely approaches the mixing chamber temperature $T_{\rm{MC}}$ (red solid line), the CBT temperature (blue data) is increased much more during ramping of the magnetic field to $B_{\rm{i}}=9\,$T and saturates significantly above $T_{\rm{MC}}$, indicating a significant heat leak onto the CBT and a weak thermal link between the CBT and the Cu plates. After precooling for almost 3 days, we obtain $T_{\rm{Cu}}\approx 10\,$mK and $T_{\rm{CBT}}\approx 24\,$mK, which sets the starting point for the nuclear demagnetization\footnote{This corresponds to a nuclear polarization of $\approx40\,\%$ in the Cu plates and $\approx17\,\%$ in the CBT islands\cite{Pobell2007}.}.

In the second AND step, shown in Fig.\,\ref{fig:cbt_AND}(b), the nuclear stage is thermally decoupled from the mixing chamber (heat switches in the superconducting state) and the magnetic field is ramped down slowly to its final value $B_{\rm{f}}=0.375\,$T. This reduces the nuclear spin temperature according to $T_{\rm{f}}=T_{\rm{i}}\cdot B_{\rm{f}}/B_{\rm{i}}$ for an ideal adiabatic process, where $T_{\rm{i}}$ and $T_{\rm{f}}$ denote initial and final temperature, respectively. The efficiency $\xi\leq 1$ of the process can be defined as the ratio of the realized and the ideal, adiabatic temperature reduction~\cite{Clark2010} $\xi=T_{\rm{i}}/T_{f} \cdot B_{f}/B_{i}$ , where $\xi=1$ corresponds to perfect adiabaticity. The AND process for the Cu plates (orange markers in Fig.\,\ref{fig:cbt_AND}(b)) is almost ideal (orange dashed line), resulting in an efficiency $\xi\gtrsim0.9$. A larger deviation is observed for the CBT (compare blue data and dashed blue line), resulting in $\xi\gtrsim0.35$. However, despite the reduced efficiency for the CBT, we obtain a significant reduction in CBT temperature by a factor of 8.6, giving a final $T_{\rm{CBT}}$ of $2.8\pm0.1\,$mK. These error bars result from the uncertainty in $E_c$, obtained from the curve fit in the inset of Fig.\,\ref{fig:differential}(b). The present measurement shows significantly improved AND efficiency compared to previous works~\cite{Bradley2016,Bradley2017} and constitutes an important step towards the $\mu$K regime.


\par To get more insight about the thermal coupling of the CBT to its environment, we monitor the warm up process after AND, shown in Fig.\,\ref{fig:cbt_AND}(c). While $T_{\rm{Cu}}$ remains almost constant during the twenty hours period of time investigated here (increase by less than $50\,\mu$K), the CBT starts to warm up immediately and reaches an equilibrium value of 7.5\,mK after eight hours. Furthermore, the CBT conductance does not recover its low temperature zero bias value when returning to zero bias after performing a bias sweep of the CBT after AND (not shown). Instead, it saturates at 7.5\,mK, while the Cu plates are still well below 1~mK. These observations indicate a rather limited spin reservoir due to the finite size of the CBT islands as well as a significant heat leak due to applied bias, in a limit where the CBT is well decoupled from the Cu plates. This indicates that the AND process directly demagnetizes the CBT while little cooling power is provided externally from the Cu plates.

\par In order to obtain a more quantitative insight, a simple thermal model is developed (schematic in inset of Fig.\,\ref{fig:cbt_AND}(a)) to qualitatively capture the main features of the experiment. For the Cu plates and the CBT islands, we have three different thermal subsystems, namely phonons, electrons and nuclei. We note that at low temperature, the largest contribution to the specific heat by far is provided by the nuclei. Electrons are coupled on one hand to the nuclear bath by the hyperfine interaction via the Korringa link~\cite{Pobell2007}, and on the other hand to the phonon bath by means of the electron-phonon interaction. Hence, nuclei and phonons are only indirectly coupled through the electronic system.

The CBT device with its large metallic islands is in principle connected thermally to the Cu plates through its electronic leads and through its substrate. For an array of resistive tunnel junctions separating the islands from the leads, the resulting Wiedemann-Franz cooling turns out to be weak. Given the insulating substrate on which the CBT resides, there is only the phonon degree of freedom available for transferring heat between Cu plates and CBT islands -- thus again giving only weak coupling. This phonon process contains several thermal resistances in series, namely the weak electron-phonon coupling in the CBT islands (negligibly weak resistance in the Cu plates due to their much larger size), the acoustic mismatch at the metal-semiconductor interfaces giving rise to a Kapitza boundary resistance~\cite{Pobell2007,Swartz1989} and finally the weak thermal conductivity within the insulating substrate itself. In presence of a finite heat leak onto the CBT islands, these weak thermal links lead to a significant temperature difference between CBT and Cu plates and associated long time constants, as observed, particularly during precooling.

Given such a limiting bottle neck between Cu plates and CBT islands, we assume the electronic, phononic and nuclear temperature within the CBT itself to be well coupled (bold arrows in the schematics) and equilibrated. Thus, in a simplified model, the CBT is taken as a single thermal system carrying the nuclear specific heat and is assumed to be weakly coupled to the Cu plate through $\dot{Q}_{\rm{coup}}=A(T_{\rm{CBT}}^p-T_{\rm{Cu}}^p)$ (dashed arrows), where the coupling constant $A$ and the exponent $p$ are fit parameters. In addition, a parasitic heat load $\dot{Q}_{\rm{par}}$ to the CBT is assumed to subsume e.g. electronic noise, heat release, or pulse tube eddy current heating\cite{SOM}.

\par We obtain qualitative agreement between model and data for all three stages of the AND process -- precool, demagnetization and warm up -- using a single coupling constant $A=7.6\cdot10^{-12}\,\rm{W/K^3}$, and a static parasitic heat leak (during precool and warm up) of $\dot{Q}_{\rm{par}}=32\,$aW per island. However, a substantially increased dynamic heat load of $\dot{Q}_{\rm{par}}=485\,$aW has to be assumed during demagnetization in order to explain the low demagnetization efficiency in Fig.\,\ref{fig:cbt_AND}(b). Furthermore, we allow for a weak temperature dependence of the coupling exponent $p$, reducing its value from $p=3.2$ during precool to $p=2.7$ during demagnetization and further down to $p=2.5$ for the warm up process. Similar exponents were also obtained in earlier works~\cite{Casparis2012,Scheller2014}. Under these assumptions, the thermal response of the CBT is qualitatively captured by the model.

\par We note that significant temperature differences between electronic and nuclear system would lead to initial cooling effects during the warm up process due the much smaller static heat leak compared to the dynamic case. This is in contrast to the measurements, thus supporting the hypothesis of equilibrated subsystems within the CBT. We also observe that the parasitic heat determined from warm up curves of the Cu plates\cite{Palma2017} is similar to that obtained from the CBT model here~\footnote{Static heat leak: $32\,$aW per CBT junction, corresponding to $5.4\,$nW/mol Cu, compared to typically $1-2$\,nW/mol Cu for the large plates. Dynamic heat leak: $485\,$aW per junction during AND, corresponding to $82\,$nW/mol Cu, compared to an estimated $30$\,nW/mol Cu for the large plates. \cite{Palma2017}}. This suggests that both the cooling power as well as the parasitic heat leak scale with the volume and area of copper used. This is confirmed with measurements on a different CBT (see supplement\cite{SOM}), showing very similar demagnetization performance despite CBT islands which are about 150 times smaller in area and volume. Thus, the predominant heat leak is not coupling in through the leads of the CBT (e.g. electronic noise or external electronic heat leak), but rather through the area or volume of the islands (e.g. microwave absorption or eddy currents and material heat release).

\par In conclusion, we demonstrate simultaneous on-and-off chip magnetic cooling of a CBT with an efficiency of $35\,\%$, thereby lowering the device temperature by a factor of 8.6 from 24\,mK down to 2.8\,mK. The CBT remains colder than the dilution refrigerator mixing chamber for more than 6 hours. Future improvements include improved microwave filtering, reduction of vibration induced eddy current heating due to active damping, better mounting and by rigidly fixing the support structure of the nuclear stage to the mixing chamber shield and magnet support assembly~\cite{todoshchenko_2014}. This should improve the currently inefficient precooling process as well as reduce the large dynamic heat leak, and thus reduce the final temperature after AND.

\begin{acknowledgments}
We would like to thank H.~J.~Barthelmess, R.~Blaauwgeers, J.~P.~Pekola, G.~Pickett, and P.~Vorselman for useful input and discussions. The work shop teams of M.~Steinacher and S.~Martin are gratefully acknowledged for technical support. This work was supported by the Swiss NSF, NCCR QSIT, the Swiss Nanoscience Institute, the European Microkelvin Platform (EMP), an ERC starting grant (DMZ), and EU-FP7 MICROKELVIN and ITN Q-NET 264034 (AVF).
\end{acknowledgments}


%

\end{document}